\documentclass[10pt,a4paper]{article}
\usepackage[margin=1.5in]{geometry}
\usepackage{amsmath,amssymb,amsthm}
\usepackage[T1]{fontenc}
\usepackage[latin1]{inputenc}
\usepackage[tiny]{titlesec}
\usepackage{hyperref}

\newcommand{\U}{\mathcal{U}}

\newcommand{\M}{\mathcal{M}}
\newcommand{\X}{\mathcal{S}}
\renewcommand{\L}{\mathcal{L}}

\newcommand{\e}{\mathfrak{e}}

\renewcommand{\d}{\mathfrak{d}_{\textsc{J}_1}}

\newcommand{\N}{\mathbb{N}}
\newcommand{\E}{\mathcal{E}}
\renewcommand{\SS}{\mathbb{S}}

\newcommand{\D}{\mathcal{D}}

\newcommand{\R}{\mathbb{R}}

\newcommand{\I}{1{\hskip -2.5 pt}\hbox{I}}
\newcommand{\II}{\mathbf{I}}

\newcommand{\ACKNO}[1]{\noindent\textbf{Acknowledgments.} #1}

\theoremstyle{plain}
\newtheorem{Thm}{Theorem}[section]
\newtheorem{Lem}[Thm]{Lemma}
\newtheorem{Prop}[Thm]{Proposition}
\newtheorem*{Prop*}{Proposition}
\newtheorem{Cor}[Thm]{Corollary}

\theoremstyle{definition}
\newtheorem{Def}[Thm]{Definition}
\newtheorem{Eg}[Thm]{Example}

\theoremstyle{remark}
\newtheorem{Rem}[Thm]{Remark}
\def \follmer {F\"{o}llmer }

\def \cadlag {c\`adl\`ag }

\def \closed {generic}

\title{A model-free approach to continuous-time finance}

\author{Henry CHIU \footnote{Dept of Mathematics, Imperial College London.
    h.chiu16@imperial.ac.uk}
  \and 
  Rama CONT\footnote{Mathematical Institute, University of Oxford.
  Rama.Cont@maths.ox.ac.uk 
  }
  }

\date{28 Novemeber 2022}
\begin{document}

\maketitle

\begin{abstract}
We present a non-probabilistic, pathwise approach to continuous-time finance based on causal functional calculus. We introduce a definition of self-financing, free from any integration concept and show that the value of a self-financing portfolio is a pathwise integral (every self-financing strategy is a gradient) and that generic domain of functional calculus is inherently arbitrage-free. We then consider the problem of hedging a path-dependent payoff across a generic set of scenarios. We apply the transition principle of Isaacs in differential games and obtain a verification theorem for the optimal solution, which is characterised by a (fully non-linear) path-dependent equation. For the Asian option, we obtain explicit solution.
\end{abstract}


\tableofcontents
\newpage
\section{Introduction}\noindent
    In an insightful expository \cite[\S 5]{FS}, \follmer and Schied sketched a non-probabilistic, pathwise  framework
    (see also \cite{BW}) as a mean to model Knightian uncertainty. In their formulation, if price paths were to evolve continuously, then they cannot be of bounded variation, since this gives rise to very simple arbitrage opportunities (free lunches), e.g.   
    
    \begin{eqnarray}\label{eq:lunch}
    (x(T)-x(0))^2-[x](T)=\int_{0}^{T}2(x(t)-x(0))dx(t),
    \end{eqnarray}where (\ref{eq:lunch}) would become non-negative for all continuous paths of bounded variation (i.e. $[x]=0$) and strictly positive for all paths meeting the condition \begin{eqnarray*}x(T)\neq x(0).\end{eqnarray*}
    
    However, if we are \emph{uncertain} that price paths would evolve continuously, then paths of bounded variations would no longer give rise to such arbitrage opportunities. Since in this case, for all \cadlag paths of bounded variation that admits at least one single discontinuity, we have
    
    \begin{eqnarray*}
    [x](T)\geq(\Delta x(t))^{2}>0,
    \end{eqnarray*}for some $t\leq T$ and it is now possible for (\ref{eq:lunch}) to go negative. In this work, we shall relax the continuity hypothesis and first investigate, rigorously, the relationship between price variation and arbitrage, that is, we do not assume price paths must possess variation a priori.
    
    We introduce the abstract formulation of (causal) functional calculus \cite{CC2} on generic path space (the minimal domain that support a functional calculus) and first obtain the analytical analogue of the (probabilistic) classical notions in mathematical finance. In particular, we introduce the definition of self-financing, free from any integration concept and show that the value of every self-financing portfolio (formed by left continuous strategies with right limit), can be represented as a pathwise integral (i.e. limit of left Riemann sums) and proved that every generic path space is arbitrage free, a fundamental property that we shall use in order to obtain the optimal hedging strategy.
    
    This arbitrage free result stands for all generic domains that include, but not limited to, paths of $p$th order variation, for any $p\in 2\N$. In contrast to related results established using the measure-theoretic, game approach of Vovk \cite{VV}, Lochowski, Perkowski \& Promel \cite{LPP} , we are able to work with the classical notion of arbitrage, rather than passing to an asymptotic relaxation that may not necessarily be implementable by a self-financing trading strategy.
    
    For non-linear payoffs, we showed that a perfect hedge does not exist in general. We adopt a primal approach to superhedging on bounded generic subset. In particular, we solve the model-free superhedging problem over the above set of scenarios using  
a minimax approach, in the spirit of Isaacs's tenet of transition \cite{RI}, and provide  a verification theorem for the optimal cost-to-go functional. As an example, we study the case of Asian options and obtain explicit solution.
    
    Related superhedging problems have been studied using probabilistic approaches or so-called robust approaches  based on quasi-sure analysis based on a family of probability measures   \cite{bartl2019,LPP,MN,NS}. In contrast to these approaches, our approach is purely pathwise and does not appeal to any probabilistic assumptions.  Also, we are able to by-pass many technical difficulties (e.g. not having to deal with duality gap and polar set). Finally, our optimal hedging strategy (delta) comes as a by-product, whereas in the quasi-sure approach, it will not be straightforward to compute the optimal delta, see for instance \cite{MN}.
    
\section{Notations}\noindent

	Denote $D$ to be the Skorokhod space of $\R^{m}$-valued positive \cadlag functions \begin{eqnarray*}t\longmapsto x(t):=(x_1(t),\ldots,x_{m}(t))'\end{eqnarray*} on $\R_{+}:=[0,\infty)$ and for $p\in 2\mathbb{N}$, we denote $D(\R_{+},\R^{m}\otimes^p)$ the Skorokgod space of $\R^{m}\otimes^p$-valued \cadlag functions on $\R_{+}:=[0,\infty)$. Denote $C$, $\SS$, $BV$ respectively, the subsets of continuous functions, step functions, locally bounded variation functions in $D$. $x(0-):=x_0>0$ and $\Delta x(t):=x(t)-x(t-)$. The path $x\in D$ stopped at $(t,x(t))$ (resp. $(t,x(t-))$)\begin{eqnarray*}s\longmapsto x(s\wedge t)
\end{eqnarray*}shall be denoted by $x_t\in D$ (resp. $x_{t-}:=x_t-\Delta x(t)\I_{[t,\infty)}\in D$). We write $(D,\d)$ when $D$ is equipped with a complete metric $\d$ which induces the Skorokhod (a.k.a. \textsc{J}$_1$) topology. 

		Let $\pi:=(\pi_n)_{n\geq 1}$ be a fixed sequence of partitions $\pi_n=(t^{n}_0,...,t^{n}_{k_n})$ of $[0,\infty)$ into intervals $0=t^{n}_0<...<t^{n}_{k_n}<\infty$; $t^{n}_{k_n}\uparrow\infty$ with vanishing mesh $|\pi_n|\downarrow 0$ on compacts. By convention, $\max(\emptyset\cap\pi_n):=0$, $\min(\emptyset\cap\pi_n):=t^{n}_{k_n}$. Since $\pi$ is fixed, we will avoid superscripting $\pi$. 
		
    For any $p\in 2\mathbb{N}$, we say that $x\in D$ has finite $p$-th order variation $[x]_{p}$ if \begin{eqnarray*}\label{eq:qv}
   \sum_{\pi_n\ni t_i\leq t}\left(x({t_{i+1}})-x({t_{i}})\right)^{\otimes p}
    \end{eqnarray*}converges to $[x]_{p}$ in the Skorokhod J$_1$ topology in $D(\R_{+},\R^{m}\otimes^p)$. In light of \cite{CC}, we remark that in the special case $p=2$, this definition is equivalent to that of \follmer \cite{HF}. We refer to \cite{CP} for a discussion of $p$-th order variation for  continuous paths. We denote $V_{p}$ the set of \cadlag paths of finite $p$-th order variations, \begin{eqnarray}t'_n:=\max\{t_i<t|t_i\in\pi_n\},\label{eq:t_n}\end{eqnarray}and the following piecewise constant approximations of $x$ by

\begin{eqnarray}
x^{n}:=\sum_{t_i\in\pi_n}x(t_{i+1})\I_{[t_i,t_{i+1})}.\label{eq:x_n}
\end{eqnarray}We let $\Omega\subset D$ be \emph{\closed}{} (Def.~\ref{def:closed}) and define our \emph{domain} as\begin{eqnarray*}\Lambda:=\{(t,x_t)|t\in\R_{+}, x\in \Omega\}.\end{eqnarray*}

\section{Casual functional calculus}\label{sec:calculus}\noindent 
Causal functional calculus \cite{CC2} is a calculus for non-anticipative functionals defined on sets of \cadlag paths satisfying certain stability properties.
     In this section, we summarize some key definitions and results; we refer to \cite{CC2} for a detailed exposition.
     
\begin{Def}[Generic sets of paths]\label{def:closed}
A non-empty subset $\Omega\subset D$ is called \emph{\closed} if $\Omega$ satisfies the following  closure properties under operations: (we recall (\ref{eq:x_n}) for the definition of $x^n$) 
	
	\begin{itemize}
	
	\item[(i)] For every $x\in \Omega$, $T> 0$, $\exists N(T)\in\N$;  $x^n \in \Omega ,\quad\forall n\geq N(T)$.
	
	\item[(ii)] For every $x\in\Omega, t\geq 0$, $\exists$ convex neighbourhood $\Delta x(t)\in \U$ of $0$; \begin{eqnarray}\label{eq:nbr}x_{t-}+e\I_{[t,\infty)}\in \Omega,\quad \forall e\in \U.\end{eqnarray}

    \end{itemize}
	\end{Def}
	
	\begin{Eg}\label{eg:closed}Examples of \closed\  subsets include
	$\SS$, $BV$, $D$ and $V_{p}$  for $p\in 2\mathbb{N}$. Generic subsets are closed under finite intersections.  All subsets of $C$ are not \closed. 
	\end{Eg}

\begin{Def}[Strictly causal functionals]\label{def:causal} Let $F: \Lambda \to \mathbb{R}$   and denote $F_{-}(t,x_t)= F(t,x_{t-})$. $F$ is called \emph{strictly causal} if $F=F_{-}$.
\end{Def}

	We associate with the sequence of partitions $\pi$ a topology on the space $\Lambda$ of \cadlag paths called the $\pi$-topology, introduced in \cite{CC2}:
\begin{Def}[Continuous functionals]\label{prop:pi}
We denote by $C(\Lambda)$ the set of maps $F:\Lambda\to \mathbb{R}$ which satisfy 
\begin{alignat*}{3}
1.&(a) \lim_{s\uparrow t; s\leq t}F(s,x_{s-})=F(t,x_{t-}),\\ 
  &(b) \lim_{s\uparrow t; s<t}F(s,x_{s})=F(t,x_{t-}),\\
  &(c)\hspace{1mm} t_n\longrightarrow t; t_n\leq t'_n \Longrightarrow F(t_n,x^{n}_{t_n-})\longrightarrow F(t,x_{t-}),\\
  &(d)\hspace{1mm} t_n\longrightarrow t; t_n<t'_n \Longrightarrow F(t_n,x^{n}_{t_n})\longrightarrow F(t,x_{t-}),\\
\newline\\
2.&(a) \lim_{s\downarrow t; s\geq t}F(s,x_{s})=F(t,x_{t}),\\
  &(b) \lim_{s\downarrow t; s>t}F(s,x_{s-})=F(t,x_{t}),\\
  &(c)\hspace{1mm} t_n\longrightarrow t; t_n\geq t'_n\Longrightarrow F(t_n,x^{n}_{t_n})\longrightarrow F(t,x_t),\\
  &(d)\hspace{1mm} t_n\longrightarrow t; t_n>t'_n\Longrightarrow F(t_n,x^{n}_{t_n-})\longrightarrow F(t,x_{t}),\\
\end{alignat*}for all $(t,x_t)\in\Lambda$. 
A functional is called \emph{left (resp. right) continuous} if it satisfies 1.(a)-(d) (resp. 2.(a)-(d)). \end{Def} 

By analogy with the concept of 'regulated functions' we define:
\begin{Def}[Regulated functionals]\label{def:version} A functional $F: \Lambda \to \mathbb{R}$ is \emph{regulated} if there exists $\widetilde{F}\in C(\Lambda)$ such that $\widetilde{F}_{-}=F_{-}$.\    $\widetilde{F}$  is  then unique by Prop.~\ref{prop:pi}.2(b).
\end{Def}
\begin{Rem}\label{rem:regulated}
Since $C(\Lambda)$ is an algebra, we remark the set of  regulated functionals forms an algebra.
\end{Rem}

\begin{Def}[Horizontal differentiability]\label{def:dt}
$F:\Lambda\longmapsto \R$ is called \emph{differentiable in time} if \begin{eqnarray*}
\D F(t,x_t):=\lim_{h\downarrow 0}\frac{F(t+h,x_t)-F(t,x_t)}{h} 
\end{eqnarray*} exists $\forall (t,x_t)\in\Lambda$.\end{Def}

\begin{Def}[Vertical differentiability]\label{def:dx}
$F:\Lambda\longmapsto \R$ is called \emph{vertically differentiable} if for every $(t,x_t)\in\Lambda$, the map 
\begin{eqnarray*}e\longmapsto F\left(t,x_t+e\I_{[t,\infty)}\right)\end{eqnarray*} is differentiable at $0$. We define $\nabla_{x}F(t,x_t):=(\nabla_{x_1}F(t,x_t),\ldots,\nabla_{x_m}F(t,x_t))'$;
\begin{eqnarray*}
\nabla_{x_i}F(t,x_t):=\lim_{\epsilon\rightarrow 0}\frac{F\left(t,x_t+\epsilon\mathbf{e}_i\I_{[t,\infty)}\right)-F(t,x_t)}{\epsilon}.
\end{eqnarray*}
\end{Def}

\begin{Def}[differentiable]
A functional is called \emph{differentiable} if it is horizontally and  vertically differentiable.
\end{Def}

\begin{Rem}
All definitions above are extended to multidimensional functions on $\Lambda$ whose components satisfy the respective conditions.
\end{Rem}

\begin{Lem}\label{Lem:causal}
A function on $\Lambda$ is strictly causal if and only if it is differentiable in space with vanishing derivative.
\end{Lem}

\begin{proof}
We refer to \cite[\S 4]{CC2}.
\end{proof}

\begin{Def}[Classes $\X$ and $\M$]\label{def:smooth}
A continuous and differentiable functional $F$ is of \emph{class $\X$} if $\D F$ is right continuous and locally bounded, $\nabla_{x}F$ is left continuous and strictly causal. If in addition, $\D F$ vanishes then $F$ is of \emph{class $\M$}. 
\end{Def}

Denote $\M(\Lambda)$ the set of all functionals of class $\M$ and $\M_{0}(\Lambda)$ the subset of $\M(\Lambda)$ with vanishing initial values. 

\begin{Def}[Pathwise integral]\label{def:pathwise}Let $\phi:\Lambda\longmapsto \R^{m}$; $\phi_{-}$ be left continuous. For every $x\in\Omega$, define \begin{eqnarray}\label{eq:discrete}\II(t,x^n_{t}):=\sum_{\pi_n\ni t_i\leq t}\phi(t_i,x^n_{t_i-})\cdot(x(t_{i+1})-x(t_i)).\end{eqnarray} If $\II(t,x_t):=\lim_{n} \II(t,x^n_{t})$ exists and $\II$ is continuous, then $\phi$ is called \emph{integrable} and $\II:=\int\phi dx$ is called the \emph{pathwise integral}.
\end{Def}

We remark that, if $\Omega\subset QV$, then integrands of the type $\nabla f\circ x, f\in C^2(\mathbb{R}^d)$ \cite{HF} and their  path-dependent analogues  $\nabla F\circ x, F\in \mathbb{C}^{1,2}(\mathbb{R}^d)$ \cite{CC2} are integrable.

\begin{Thm}[Representation theorem]\label{thm:ftc} 

A functional $F:\Lambda \to \mathbb{R}$ is a pathwise integral if and only if  $F\in \M_{0}(\Lambda)$:
$$F\in  \M_{0}(\Lambda) \iff \exists \phi:\Lambda\to \R^{m}, \phi_{-} \text{ left-continuous};$$

$$
F(t,x_t)=\int_0^t\phi_{-} dx, \quad\forall (t,x_t)\in \Lambda.$$\end{Thm}

\begin{proof}
We refer to \cite[\S5]{CC2}.
\end{proof}

\section{Market scenarios,  self-financing strategies and  arbitrage}\label{sec:trading}\noindent 

    We consider a frictionless market with $d>0$ tradable assets, and one numeraire whose price is identically 1. We denote $x$ to be the price paths of tradable assets and $x\in\Omega$, where $\Omega$ is generic Def.~\ref{def:closed}.
    
    The number of shares in assets $\phi$ and the numeraire $\psi$ held immediately before the portfolio revision at time $t$ will be denoted by $\phi_{-}$ and $\psi_{-}$.
    
    We aim to address the following fundamental questions:
    \begin{itemize}
        \item What is \emph{self-financing}? Since there is not a priori that the value of a portfolio $V$ must be expressible as $dV=\phi dx$.
        \item What is \emph{no-arbitrage}? Is it necessary that price paths must possess variation of some sort? 
    \end{itemize}
    
    A trading strategy is a pair $(\phi,\psi)$ of regulated functionals   $\phi: \Lambda \mapsto \R^{d}$ and $\psi: \Lambda \mapsto \R$. 
    The value $V$ of the portfolio is given by \begin{eqnarray}
V(t,x_t):=\widetilde{\phi}(t,x_{t})\cdot x(t)+\widetilde{\psi}(t,x_{t}).\label{eq:self2}
\end{eqnarray}
  A key concept is the concept of \emph{self-financing} strategy \cite[\S2]{BW}.
  This concept is usually defined in a probabilistic setting, by equating the changes in the portfolio value $V$ with a {\it gain process} defined as a stochastic integral $\int \phi_-.dS$. The situation in a non-probabilistic setting is subtle. There exists different approaches to the definition of a "pathwise integral" in model-free mathematical finance, which may lead to different notions of "self-financing", i.e. the \emph{pitfall} that if we can express $dV=\phi dx$, then the portfolio would be called "self-financing"! Consequently, this has also led to different results on arbitrage.
  
  We propose a new approach to this concept based on  {\it local} properties, without involving any use of  (pathwise or stochastic) integration notions:
\begin{Def}[Self-financing strategy]\label{def:self}\noindent\\
A trading strategy or portfolio $(\phi,\psi)$ is called \emph{self-financing} if for every $(t,x)\in \Lambda$,
\begin{itemize}
    \item [(i)] $\Delta \widetilde{\phi}(t,x_t)\cdot x(t)+\Delta\widetilde{\psi}(t,x_t)=0$,
    \item [(ii)] $\left(\widetilde{\phi}(t+h,x_t)-\widetilde{\phi}(t,x_t)\right)\cdot x(t)+\widetilde{\psi}(t+h,x_t)-\widetilde{\psi}(t,x_t)=0$ for all $h>0$. 
\end{itemize}
\end{Def}
Both conditions correspond to the property that the proceeds from any change in the asset position is reflected in the change in the cash position. However the important point is that we only require this in two situations: \begin{itemize}
    \item[(i)] an instantaneous change in the asset position, and 
    \item[(ii)] a change in the asset/cash position while the asset prices remain constant.
\end{itemize}
As we will show, through piecewise constant approximation these two situations cover the case of all continuous-time strategies under minimal regularity properties.

\begin{Rem}
If $(\phi,\psi)$ is self-financing, then the value of the portfolio may also be expressed as\begin{eqnarray}
V(t,x_t)={\phi}(t,x_{t-})\cdot x(t)+{\psi}(t,x_{t-}).\label{eq:self1}
\end{eqnarray}We remark here that interchanging  (\ref{eq:self2}) and (\ref{eq:self1}) for the definition of a portfolio value would not have any effect for self-financing portfolios.  
\end{Rem}

\begin{Thm}[Gain of a self-financing strategy as a pathwise integral]\label{thm:self}
Let $V$ be the portfolio value associated with the trading strategy $(\phi,\psi)$. Then  $(\phi,\psi)$ is self-financing if and only if 
   $V\in\M(\Lambda)$, $\nabla_{x}V=\phi_{-}$. In that case \begin{eqnarray}
     V(t,x_t)=V(0,x_0)+\int_{0}^{t}\phi(s,x_{s-})dx.\label{eq:gain}
    \end{eqnarray}
\end{Thm}

\begin{proof}
If $(\phi,\psi)$ is self-financing, we may first use (\ref{eq:self1}) to deduce that $\nabla_{x}V=\phi_{-}$, which is left continuous and strictly causal. From (\ref{eq:self2}) and the fact that $C(\Lambda)$ is an algebra (i.e. Prop.~\ref{prop:pi}), we see that $V$ is continuous. We then apply (\ref{eq:self2}) and Def.~\ref{def:self}(ii) to deduce that $\D V$ is vanishing. Hence, $V\in\M(\Lambda)$ and (\ref{eq:gain}) follows from Thm.~\ref{thm:ftc}. 

On the other hand, if $V\in\M(\Lambda)$, then $V$ is continuous. By the continuity of $V$, (\ref{eq:self1}) and Prop.~\ref{prop:pi}.2(b), we first obtain (\ref{eq:self2}), hence Def.~\ref{def:self}(i). Since $\D V$ vanishes, by \cite[Lem.5.1]{CC2}, we obtain \begin{eqnarray*}V(t+h,x_t)-V(t,x_t)=\int_{t}^{t+h}\D V(s,x_t)ds=0.\end{eqnarray*}Resorting once again to (\ref{eq:self2}), we also obtain Def.~\ref{def:self}.(ii), hence $(\phi,\psi)$ is self-financing.
\end{proof}

\begin{Prop}\label{prop:self}
Let $V\in\M(\Lambda)$, then the following properties are equivalent:
\begin{itemize}
    \item [(i)] $V$ is the  value of a self-financing trading strategy $(\phi,\psi)$.
    \item [(ii)] $\nabla_{x}V$ is regulated.
\end{itemize}

\end{Prop}

\begin{proof}
(i) implies (ii) follows from Def.~\ref{def:self}, Def.~\ref{def:version} and Thm.~\ref{thm:self}. Assume (ii) holds, let $\phi$ be the continuous version of $\nabla_{x}V$ and put \begin{eqnarray}\psi(t,x_t):=V(t,x_t)-\phi(t,x_t)\cdot x(t),\label{eq:self3}\end{eqnarray}then $\psi$ is continuous (i.e. $C(\Lambda)$ is an algebra) and $V$ is the portfolio value associated with the trading strategy $(\phi,\psi)$. Taking $\Delta$ from (\ref{eq:self3}), we obtain \begin{eqnarray}
\Delta V-\nabla_{x}V\Delta x=x\cdot \Delta\phi+\Delta\psi.\label{eq:self4}
\end{eqnarray}By Thm.~\ref{thm:ftc}, we deduce the LHS of (\ref{eq:self4}) vanishes and obtain (\ref{eq:self1}), hence $\nabla_{x}V=\phi_{-}$, the proof is complete by Thm.~\ref{thm:self}. 
\end{proof}

\begin{Rem}[Self-financing $V$]\label{rem:self}
In view of Thm.~\ref{thm:self}, Prop.~\ref{prop:self} \& (\ref{eq:self3}), we may call a functional $V$  \emph{self-financing} if $V\in\M$ with regulated $\nabla_{x}V$. 
\end{Rem}

\begin{Def}[Arbitrage]\label{def:arbitrage}\noindent\\
 A self-financing strategy $(\phi,\psi)$ with value $V$ is called an \emph{arbitrage} on $[0,T]$ if \begin{eqnarray}
\forall x\in\Omega,\qquad  V(T,x_{T})-V(0,x_{0})\geq 0\label{eq:arbitrage}
 \end{eqnarray} and there exists $x\in\Omega$ such that $V(T,x_{T})-V(0,x_{0})>0$.
\end{Def}

 \begin{Lem}\label{lem:fair}
 Let $M\in\M_{0}(\Lambda)$. If there exists $T>0$; \begin{eqnarray*}M(T,x_{T})\geq 0\end{eqnarray*}$\forall x\in\Omega$,
 then for every $x\in \Omega$, the map \begin{eqnarray*}
 t\longmapsto M(t,x_{t})
 \end{eqnarray*}is non-negative for $ t\leq T$.
 \end{Lem}
 
 \begin{proof}
If $M\in\M_{0}(\Lambda)$, then $\D M$ vanishes, by \cite[Lem.5.1]{CC2} we obtain\begin{eqnarray*}
M(t,x_{t})=M(t,x_{t})+\int_{t}^{T}\D M(s,x_t)ds= M(T,x_{t})\geq 0
\end{eqnarray*}for all $t\leq T$, where the last inequality is due to $x_t\in\Omega$.
 \end{proof}

 \begin{Thm}[Fair game]\label{thm:fair}
 Let $M\in\M_{0}(\Lambda)$. If there exists $T>0$; \begin{eqnarray*}M(T,x_{T})\geq 0\end{eqnarray*} for all $x\in\Omega$, then $M(T,x_{T})\equiv 0$.
 \end{Thm}
 
\begin{proof}
Let $T>0$; $M(T,x_{T})\geq 0$ $\forall$ $x\in \Omega$. By Lem.~\ref{lem:fair}, we first obtain \begin{eqnarray}
M(t,x_{t})\geq 0\label{eq:non-negativ}
\end{eqnarray}for all $t\leq T$, $x\in \Omega$. Suppose there exists $\omega\in\Omega$; \begin{eqnarray*}M(T,\omega_T)>0.\end{eqnarray*}By the continuity of $M$ and Thm.~\ref{thm:ftc}, it follows \begin{eqnarray}M(T,\omega^{n}_T)=\sum_{\pi_n\ni t_i\leq T}\nabla_{x}{M}(t_i,\omega^n_{t_i-})(\omega(t_{i+1})-\omega(t_i))>0\label{eq:martingale}\end{eqnarray}for all $n$ sufficiently large. Define \begin{eqnarray*}t^{*}_n:=\min\{t_i\in\pi_n | M(t_i,\omega^{n}_{t_i})>0\},\end{eqnarray*}then $t^{*}_n\leq T$. By (\ref{eq:non-negativ}), (\ref{eq:martingale}), the left continuity of $M$ and the fact that $\omega^{n}\in\Omega$, we obtain  \begin{eqnarray*}
M(t^{*}_n,\omega^{n}_{t^{*}_n})>M(t^{*}_n,\omega^{n}_{t^{*}_n-})=0,
\end{eqnarray*}hence
\begin{alignat*}{2}
M(t^{*}_n,\omega^{n}_{t^{*}_n})=\nabla_{x}{M}(t^{*}_n,\omega^n_{t^{*}_n-})\Delta \omega(t^{*}_n)>0.
\end{alignat*}Def.~\ref{def:closed}(ii) implies that there exists $\epsilon>0$; \begin{eqnarray*} \omega^{*}:=\omega^{n}_{t^{*}_n-}-\epsilon\Delta \omega(t^{*}_n)\I_{[t^{*}_n,\infty)}\in\Omega,\end{eqnarray*}hence 
\begin{alignat*}{2}
M(t^{*}_n,\omega^{*}_{t^{*}_n})=\nabla_{x}{M}(t^{*}_n,\omega^n_{t^{*}_n-})(-\epsilon\Delta \omega(t^{*}_n))<0,
\end{alignat*}which is a contradiction to (\ref{eq:non-negativ}).
\end{proof}
 
Using these  results we can now show that  if the set of market scenarios is a generic set of paths, arbitrage  in the sense of Def. \ref{def:closed} may not exist:
 \begin{Cor}\label{cor:arbitrage} 
 Arbitrage does not exist in a generic market.
 \end{Cor}
 
 \begin{proof}
It is a direct consequence of Def.~\ref{def:arbitrage}, Thm.~\ref{thm:self} and Thm.~\ref{thm:fair}. 
 \end{proof}
 
 \begin{Rem} As previously discussed,  the set $\SS$ of piecewise-constant paths, the space $D([0,\infty),\mathbb{R}_+^m)$ of positive \cadlag paths or the space $V_{p}$ of \cadlag paths with finite p-th order variation for $p\in 2\mathbb{N}$ are examples of generic sets of paths, to which the above result applies.
 However, unlike the results of \cite{LPP,schied2016,VV} the proof of the above result does not involve any assumption on the variation index of the path.
 \end{Rem}
 
 \section{When does a payoff admit a perfect hedge?}\label{sec:payoff}\noindent 
 
    In this section, we first uncover, from examples such as Asian, lookback and passport  options, the mathematical properties of a \emph{payoff} in the context of functional calculus.
    
    We define what a (non-) \emph{linear} payoff is and prove that a payoff can be perfectly hedged in a generic market if and only if it is linear such as the Asian option with 0 strike. 
    
   For $u, v\in \R^{l}$, we write $u>v$ if $u_i>v_i$ for all $i$. We call $v$ positive if $v>0$. Let $\Omega$ be a  generic set of paths. In order for the operation \begin{eqnarray*}x_{t-}+\e(t,x_{t-})\I_{[t,\infty)}\in\Omega\end{eqnarray*}to be closed, $\e(t,x_{t-})$ may not take arbitrary values, this motivates the following
  
\begin{Def}[admissible perturbation]\label{def:regular} A regulated function
$\e:\Lambda\to \R^{d}$ is called an \emph{admissible} perturbation if for every $x\in\Omega$, $t\geq 0$, \begin{eqnarray*}
x_{t-}+\e(t,x_{t-})\I_{[t,\infty)}\in\Omega
\end{eqnarray*}We denote $\E$ to be the set of all admissible perturbations on $\Lambda$. 
\end{Def}

\begin{Eg}\label{eg:regular}
$\e:=0$ is admissible. If $\Omega$ is either $\SS$, $V_{p}$ or $D$, then every $\R^{d}$-valued regulated function $\e$ satisfying \begin{eqnarray*}\e(t,x_{t-})>-x(t-),\end{eqnarray*}for all $x\in\Omega$, $t\geq 0$ is admissible.
\end{Eg}

\begin{Def}[non-degenerate]\label{def:degen}
A subset $\Omega$ is called \emph{non-degenerate}, if there exists $\e^{1},\ldots,\e^{d}\in\E$ where\begin{eqnarray} 
\e^{i}_j(t,x_{t-})
\begin{cases}
    \neq 0, & \text{if $i=j$}.\\
    =0, & \text{otherwise};
  \end{cases}\label{eq:degen}\end{eqnarray}for every $x\in\Omega$, $t\geq 0$.
\end{Def}

\begin{Rem}\label{rem:non-degenerate}
If $\Omega$ is either $\SS$, $V_{p}$ or $D$, then $\Omega$ is non-degenerate. In the sequel, we shall assume that $\Omega$ is non-degenerate.
\end{Rem}

\begin{Def}[Payoff]\label{def:payoff}\noindent\\
A \emph{payoff} with maturity $T>0$ is a functional $H:\Omega\to \mathbb{R}$ such that
\begin{itemize}
    \item [(i)] $H(x)=H(x_T)$ for all $x\in\Omega$.\\
    \item [(ii)] For every $x\in\Omega$, $t\geq 0$ and the map \begin{eqnarray*}e\longmapsto H(x_{t-}+e\I_{[t,\infty)})\end{eqnarray*} is continuous on every convex neighborhood $\U\subset\R^{d}$ of $0$ \newline satisfying (\ref{eq:nbr}).
\item [(iii)] The functional $(t,x_t)\longmapsto H(x_{t}) $ is continuous on $\Lambda$ and for every $\e\in\E$, the functional
\begin{eqnarray*}
(t,x_t)\in \Lambda\longmapsto H(x_{t-}+\e(t,x_{t-})\I_{[t,\infty)}),
\end{eqnarray*}is regulated.
\end{itemize}
\end{Def}
 
 
\begin{Eg}\label{eg:payoff}
Let $d=1$, $T>0$, $K\geq 0$ and let $V$ be the value of a self-financing portfolio. Then 
\begin{itemize}
\item [(a)] $H(x):=\left(\frac{1}{T}\int_{0}^{T}x(t)dt-K\right)^{+}$,
\item [(b)] $H(x):=\left(\sup_{s\leq T}x(s)-x(T)\right)^{+}$,
\item [(c)] $H(x):=\left(V(T,x_T)-K\right)^{+}$, 
\end{itemize} satisfy Definition \ref{def:payoff}.
\end{Eg}
 
\begin{proof}
We first compute $H(x_{t-}+ e\I_{[t,\infty)})$ and obtained the followings:
\begin{alignat*}{3}
&(a)\quad && \left(\frac{1}{T}\left(\int_{0}^{t\wedge T}xds+(T-t)(x(t-)+e)\I_{[0,T]}\right)-K\right)^{+},\\
&(b)\quad && \left(\sup_{s<t}x_{T}(s)-x_{T}(t-)- e\I_{[0,T]}\right)^{+},\\
&(c)\quad && \left(V(t,x_{(t\wedge T)-})+\nabla_{x}V(t,x_{t-})e\I_{[0,T]}-K\right)^{+},
\end{alignat*}which are all continuous in $e$ and we obtain Def.~\ref{def:payoff}(ii). If we replace $e$ with $\Delta x(t)$ and observe in (b) that \begin{eqnarray*}\left(\sup_{s< t}\ x_{T}(s)-x_{T}(t)\right)^{+}=\left(\sup_{s\leq t}\ x_{T}(s)-x_{T}(t)\right)^{+},
\end{eqnarray*}we see that $(t,x_t)\mapsto H(x_{t})$ is continuous. If we replace $e$ with $\e\in\E$, by the admissibility of $\e$ and Rem.~\ref{rem:regulated}, we obtain Def.~\ref{def:payoff}(iii).
\end{proof}
 
\begin{Def}[Vertically affine functionals]\label{def:linear}
A payoff $H:\Omega\to \mathbb{R}$ is called \emph{vertically affine} if for every $x\in\Omega$, $t\geq 0$ and convex neighborhood $\U\subset\R^{d}$ of $0$ satisfying (\ref{eq:nbr}), the map
\begin{eqnarray*}
e\longmapsto H(x_{t-}+e\I_{[t,\infty)}) 
\end{eqnarray*} is affine  on $\U$.
\end{Def}

\begin{Rem}
If $K=0$, the payoffs in Example~\ref{eg:payoff}(i)\&(iii) are vertically affine.
\end{Rem}

\begin{Def}[Perfect hedge]
A payoff $H:\Omega\to \mathbb{R}$   with maturity $T>0$ is said to admit a \emph{perfect hedge} on $\Omega$ if there exists a self-financing portfolio with value  $V$ such that
\begin{eqnarray*}
\forall x \in \Omega,\qquad  V(T,x_T)=H(x_T).
\end{eqnarray*}
\end{Def}

\begin{Thm}\label{thm:perfect}
Every vertically affine payoff admits a perfect hedge on $\Omega$.
\end{Thm}
\begin{proof}
If $H$ is vertically affine, then $e\longmapsto H(x_{t-}+e\I_{[t,\infty)})$ is an affine map. Since $\Omega$ is generic, it follows there exists a convex neighborhood $\Delta x(t)\in\U\subset \R^{d}$ of $0$ satisfying (\ref{eq:nbr}) and we obtain a constant $c$ and a $\phi\in\R^{d}$;\begin{eqnarray}
H(x_{t-}+e\I_{[t,\infty)})=c(t,x_{t-})+\phi(t,x_{t-})\cdot e,\label{eq:affine}
\end{eqnarray}on $\U$, hence $c(t,x_{t-})=H(x_{t-})$ and \begin{eqnarray}
H(x_{t})-H(x_{t-})=\phi(t,x_{t-})\cdot \Delta x(t).\label{eq:affine2} 
\end{eqnarray}Since it holds for every $x\in\Omega$ and $t\geq 0$, it follows from Def.~\ref{def:payoff}(iii) that \begin{eqnarray*}
V(t,x_t):=H(x_t),
\end{eqnarray*}is continuous on $\Lambda$, $\D V$ vanishes and by (\ref{eq:affine2}) and Lem.~\ref{Lem:causal}, $\nabla_{x}V(t,x_t)=\phi(t,x_{t-})$ which is strictly causal and $V$ is of class $\M$. It remains to show that $\phi$ is regulated. 
Since $\Omega$ is non-degenerate, there exists everywhere non-vanishing $\e^{i}\in\E$, $i=1,\ldots,d$;
\begin{eqnarray}\label{eq.hedge}
H(x_{t-}+\e^{i}(t,x_{t-})\I_{[t,\infty)})-H(x_{t-})=\phi(t,x_{t-})\cdot \e^{i}(t,x_{t-}).
\end{eqnarray} 
Since $(\e^{i}_i)\neq 0$, it follows from Def.~\ref{def:payoff}(iii), Rem.~\ref{rem:regulated} and \eqref{eq.hedge} that $\phi$ is regulated. By Prop.~\ref{prop:self} \& Rem.~\ref{rem:self}, $V$ is self-financing and hence the claim follows.
\end{proof}

\begin{Cor}\label{cor:perfect}
A payoff admits a perfect hedge on a generic set of paths $\Omega$ if and only if it is vertically affine.
\end{Cor}

\begin{proof}
The if part follows from Thm.~\ref{thm:perfect}. If $H$ admits a perfect hedge then there exists $V\in\M(\Lambda)$; $H(x_T)=V(T,x_T)$ on $\Omega$. It follows that \begin{eqnarray*}H(t,x_{t-}+e\I_{[t,\infty)})=V(t,x_{t-})+\nabla_{x}V(t,x_{t-})e.\end{eqnarray*}
\end{proof}

\begin{Eg}[Asian with $K=0$]\label{eg:asian}
The Asian payoff with $K=0$ i.e. Eg.~\ref{eg:payoff}(i) is linear and the perfect hedge is computed as:\begin{alignat}{2}
\nabla_{x}V(t,x_{t})&=\frac{T-t}{T},\nonumber\\
V(t,x_t)&=\frac{1}{T}\left(\int_{0}^{t\wedge T}x(s)ds+(T-t)x(t)\right),\label{eq:nullk}\\
V(0,x_0)&=x(0).\nonumber
\end{alignat}We remark here that the perfect hedge is model independent.
\end{Eg}

 \section{Hedging strategy for non-linear payoffs: Asian option}\label{sec:payoff2}\noindent 

    In the last section, we have established a very important fact, i.e. that a perfect hedge does not exist for non-linear payoffs, thereby justifying the search for an alternative approach. A well-studied paradigm for valuation in the absence of perfect replicating strategies is super-hedging across a scenarios of paths \cite{AP,TL}. 
    A well-known defeat of the superhedging approach is that the price would, in general, be too high. Here, we consider to hedge on a \emph{bounded} generic set of paths. This approach is model-free, realistic (e.g. price paths needed not be continuous) and gives a reasonable price that corresponds to the range of bounds.
    
    Let $\Omega$ be generic. We define, for fixed constants $0\leq a<b$,
\begin{eqnarray*}
\Omega_{a}^{b}:=\{x\in\Omega|a<x(t)<b\}.\end{eqnarray*} Observe that $\Omega_{a}^{b}$ is again generic, hence is itself free of arbitrage in the sense of Def. \ref{def:arbitrage}. Also, the superhedging price obtained on this set is a proper arbitrage-free price from the standpoint of $\Omega$. We denote \begin{eqnarray*}
   {\Omega}_{a}^{b}(x_t):=\{z\in{\Omega}_{a}^{b}|z_t=x_t \},\qquad \L:=\{\nabla_{x}V| V \text{ is self-financing}\}.\end{eqnarray*}


    \begin{Def}[Superhedging price and strategy]\label{def:strategy}\noindent\\
Let $H$ be a payoff defined on $\Omega$ with maturity $T>0$ and $V$ be self-financing (Rem.~\ref{rem:self})   that dominates $H$ on $\Omega_{a}^{b}$, i.e.
\begin{eqnarray}
\qquad V(T,x_T)\geq H(x_T),\label{eq:hedge}
\end{eqnarray}for all $x\in\Omega_{a}^{b}$. If for every other self-financing $W$ that dominates $H$ on $\Omega_{a}^{b}$, we have \begin{eqnarray} W(0,x_0)\geq V(0,x_0),\label{eq:hedge2}\end{eqnarray} then $V(0,x_0)$ is  called the superhedging price of $H$ and $\nabla_{x}V$ is a superhedging strategy for the payoff $H$.
\end{Def}

A superhedging strategy, if it exists, may not be unique. We first develop the notion of optimal strategy (if exists, will be unique), in the spirit of Isaacs's tenet of transition in differential games \cite[p3]{RI}. 
Our approach here is to construct a (cost-to-go) functional $U\in\X$ (Def.~\ref{def:smooth}) such that for all $0\leq s\leq t\leq T$ and $x\in\Omega_{a}^{b}$, the followings hold:
\begin{alignat}{2} U(s,x_{s})&=\min_{\phi\in\L}\sup_{z\in\Omega_{a}^{b}(x_s)}\left\{U(t,z_t)-\int_{s}^{t}\phi dz\right\},\label{eq:issacs}\\
U(T,x_T)&=H(x_T)\nonumber.
\end{alignat}
\begin{Lem}\label{lem:super}
Let $U\in\X$  be a functional that satisfies (\ref{eq:issacs}). Then 
the map \begin{eqnarray*}
h\longmapsto U(s+h,x_s) 
\end{eqnarray*}is monotonic decreasing in $[0,\infty)$.
\end{Lem}

\begin{proof}
We have \begin{eqnarray*} U(s,x_{s})\geq\min_{\phi\in\L}\left\{U(t,z_t)-\int_{s}^{t}\phi dz\right\}
\end{eqnarray*}for all $z\in\Omega_{a}^{b}(x_s)$, this holds, in particular for all $z$ stopped at $s$. It follows \begin{eqnarray*}
U(s,x_{s})\geq\min_{\phi\in\L}U(t,z_s)=U(t,x_s).
\end{eqnarray*}
\end{proof}

\begin{Rem}[The hedging portfolio]\label{rem:pnl}\noindent\\
Thus if  $U$ satisfies (\ref{eq:issacs}), then $V(T,x_T):=U_0+\int_0^{T} \nabla_{x}U dx$ solves (\ref{eq:hedge}), meeting condition (\ref{eq:hedge2}) and the solution is unique up to $\Omega_{a}^{b}$ due to
\begin{eqnarray*}
U_1(t,x_{t})=\min_{\phi\in\L}\sup_{z\in{\Omega}_{a}^{b}(x_t)}\left\{H(T,z_T)-\int_{t}^{T}\phi dz\right\}=U_2(t,x_t).
\end{eqnarray*}

In particular, $U(t_0,x_{t_0})$ is the superhedging price to hedge starting at time $0\leq t_0<T$. The relationship with the value of the hedging portfolio $V$ (See also Rem.~\ref{rem:self}) is
\begin{eqnarray*}
V(t,x_t)=U(t_0,x_{t_0})+\int_{t_0}^{t}\nabla_{x}Udx=U(t,x_t)-\int_{t_0}^{t}\D Uds, 
\end{eqnarray*}hence at maturity time $T$, the final portfolio value is\begin{eqnarray*}
V(T,x_T):=H(T,x_T)-\int_{t_0}^{T}\D Uds\geq H(T,x_T),
\end{eqnarray*}where the last inequality is due to Lemma \ref{lem:super} and the final PnL is $\int_{t_o}^{T} -\D U ds$.
\end{Rem}

We now use the following Minimax Theorem to prove a verification theorem.

\begin{Thm}[Minimax]\label{lem:minimax}
If $M\in\M$; then \begin{eqnarray*} \min_{\phi\in\L}\max_{z\in{\Omega}_{a}^{b}(x_s)}\left\{\int_{s}^{t}(\nabla_{x}M-\phi) dz\right\}=0=\max_{z\in{\Omega}_{a}^{b}(x_s)}\min_{\phi\in\L}\left\{\int_{s}^{t}(\nabla_{x}M-\phi) dz\right\}
\end{eqnarray*}
\end{Thm}

\begin{proof}
We first have 
\begin{alignat*}{2} c:=&\inf_{\phi\in\L}\sup_{z\in{\Omega}_{a}^{b}(x_s)}\left\{\int_{s}^{t}(\nabla_{x}M-\phi) dz\right\}\\
\leq&\max_{z\in{\Omega}_{a}^{b}(x_s)}\left\{\int_{s}^{t}(\nabla_{x}M-\nabla_{x}M) dz\right\}=0.
\end{alignat*}If $c<0$, then there exists an $\epsilon>0$ such that\begin{eqnarray*}
\int_{s}^{t}(\phi_{\epsilon}-\nabla_{x}M) dz\geq -(c+\epsilon)>0,
\end{eqnarray*}which gives an arbitrage. It follows from Theorem \ref{thm:fair} that $c=0$ and hence the infimum and supremum are attained respectively by $\phi:=\nabla_x M$ and any $z$. The case of maximin follows similar lines of proof.
\end{proof}

We obtain as a corollary, yet another look at functionals of class $\M$, in reminiscent to their probabilistic counterparts:

\begin{Cor}
Define for $H:\Lambda\longmapsto\mathbb{R}$

\begin{equation*}
    \mathbb{E}(H(t,x_t)|x_s):=
    \begin{cases}
      \underset{\phi\in\L}{\min}\underset{z\in\Omega_{a}^{b}(x_s)}{\sup}\left\{H(t,z_t)-\int_{s}^{t}\phi dz\right\}, & \text{if RHS exists}\\
      \infty, & \text{otherwise}.
    \end{cases}
  \end{equation*}Then for every $M\in\M(\Lambda)$, we have
$$ \mathbb{E}\left(M(t,x_t)|x_s\right)=M(s,x_s).
$$
\end{Cor}

\begin{Thm}[Verification theorem]\label{thm:veri}
Let $U\in \X(\Lambda), \nabla_{x}U\in \L$; $U$ satisfies \begin{alignat}{2}\label{eq:veri}
\sup_{z\in{\Omega}_{a}^{b}(x_t)}\int_{t}^{T}\D U(s,z_{s})ds&=0,\\
U(T,x_T)&=H(x_T),\nonumber
\end{alignat}for all $t \leq T$ and $x\in\Omega_{a}^{b}$. Then $\phi:=\nabla_{x}U$ is the superhedging strategy for $H$ on ${\Omega}_{a}^{b}$ and achieves the optimum in \eqref{eq:issacs}.
\end{Thm}

\begin{proof}

We first obtain
\begin{alignat*}{2} c:&=\inf_{\phi\in\L}\sup_{z\in{\Omega}_{a}^{b}(x_s)}\left\{\int_s^t\D U(r,z_r)dr+\int_{s}^{t}(\nabla_{x}U-\phi) dz\right\}\\
&\leq \min_{\phi\in\L}\max_{z\in{\Omega}_{a}^{b}(x_s)}\left\{\int_{s}^{t}(\nabla_{x}U-\phi) dz\right\}=0, 
\end{alignat*}due to Lem.\ref{lem:super} \& Thm.\ref{lem:minimax} (Minimax). It remains to show that $c\geq 0$.
\begin{alignat*}{2} c&\geq\sup_{z\in{\Omega}_{a}^{b}(x_s)}\inf_{\phi\in\L}\left\{\int_s^t\D U(r,z_r)dr+\int_{s}^{t}(\nabla_{x}U-\phi) dz\right\}\\
&\geq \sup_{z\in{\Omega}_{a}^{b}(x_s)}\left\{\int_s^t\D U(r,z_r)dr\right\}+\max_{z\in{\Omega}_{a}^{b}(x_s)}\min_{\phi\in\L}\left\{\int_{s}^{t}(\nabla_{x}U-\phi) dz\right\}=0,
\end{alignat*}by (\ref{eq:veri}) and Thm. \ref{lem:minimax} (Minimax). The infimum is attained by $\phi:=\nabla_xU$.
\end{proof}

\begin{Eg}[Asian option]
Let $\Omega$ be either $BV$, $V_p$; $p\in 2\mathbb{N}$ or $D$. The optimal cost-to-go functional is \begin{alignat}{2}\label{eq:asian}
U(t,x_t)=H^{+}(t,x_t)p(x(t))+H^{-}(t,x_t)(1-p(x(t))
\end{alignat}where 

\begin{alignat*}{2}
H^{+}(t,x_t)&=\left(\frac{1}{T}\left(\int_{0}^{t}x(s)ds+b(T-t)\right)-K\right)^{+},\\
H^{-}(t,x_t)&=\left(\frac{1}{T}\left(\int_{0}^{t}x(s)ds+a(T-t)\right)-K\right)^{+},\\
p(x(t))&=\frac{x(t)-a}{b-a},
\end{alignat*}and the optimal strategy is

\begin{eqnarray*}
\nabla_{x}U(t,x_t)=\frac{H^{+}(t,x_t)-H^{-}(t,x_t)}{b-a}.
\end{eqnarray*}
\end{Eg}

\begin{proof}
We first see that $U$ is of class $\X$ with $U(T,x_T)=H(x_T)$. For $z\in\Omega_{a}^{b}(x_t)$, we have \begin{eqnarray*}
\D U(s,z_s)=\D H^+(s,z_s)p(z) + \D H^-(s,z_s)(1-p(z)),
\end{eqnarray*}where

\begin{alignat*}{2}
\D H^{+}(s,z_s)&=\frac{z(s)-b}{T}\I_{\{H^+>0\}},\\
\D H^{-}(s,z_s)&=\frac{z(s)-a}{T}\I_{\{H^->0\}}.
\end{alignat*}Since $H^{+} = 0$ implies $H^{-}=0$ and that $H^{-}>0$ implies $H^{+}>0$, it follows\begin{eqnarray*}
\D U(s,z_s)=\frac{(z(s)-b)(z(s)-a)}{T(b-a)}\I_{\{H^+>0\}}\I_{\{H^-=0\}}\leq 0.
\end{eqnarray*}For sufficiently small $\epsilon>0$, we construct a path $z^{\epsilon}\in\Omega_{a}^{b}(x_t)$:

\begin{eqnarray*}
    z^{\epsilon}(s):=
    \begin{cases}
      x(t), & s\in[t,t+\epsilon)\\
      b-\epsilon, & [t+\epsilon,\infty),
    \end{cases}
  \end{eqnarray*}and observe that 
\begin{eqnarray*}
0\geq\int_{t}^{T}\D U(s,z^{\epsilon}_{s})ds\geq -\epsilon\left(1-\frac{\epsilon}{b-a}\right),
\end{eqnarray*}hence $\underset{z\in{\Omega}_{a}^{b}(x_t)}{sup}\int_t^{T}\D U(s,z_s)ds=0$ and we obtained (\ref{eq:veri}) in Thm.\ref{thm:veri}.
\end{proof}

\begin{Rem}
Note that if $K=0$, we obtain the perfect hedge in Example \ref{eg:asian} (\ref{eq:nullk}) as a special case. If we set $a=0$ and let $b\uparrow\infty$, then (\ref{eq:asian})  converges to the superhedging price (whole space) of the Asian option\begin{eqnarray*}
U(t,x_t)=\left(\frac{1}{T}\int_0^{t}x(s)ds-K\right)^{+}+x(t)\frac{T-t}{T}.
\end{eqnarray*}
\end{Rem}

\ACKNO{This research was supported by the UKRI-EPSRC under project reference 1824430 "Analysis and control of path-dependent random systems".}

\bibliographystyle{MF}
\bibliography{MF}

\end{document}